\documentclass[10pt, conference, letterpaper]{IEEEtran}

\usepackage{graphicx}
\usepackage{color}
\usepackage{url}
\usepackage[cmex10]{amsmath}
\usepackage{verbatim}
\usepackage[T1]{fontenc}
\usepackage{longtable}
\usepackage{tabularx}
\usepackage{array}
\usepackage{cleveref}
\usepackage{multirow}
\usepackage{color}
\usepackage{enumitem}
\usepackage[table]{xcolor}

\title{rTraceroute: R\'eunion Traceroute Visualization}
\author{\IEEEauthorblockN{
Xavier Nicolay\IEEEauthorrefmark{1},
R\'ehan Noordally\IEEEauthorrefmark{1} and
Yassine Gangat\IEEEauthorrefmark{1}\IEEEauthorrefmark{2}}
\\

\IEEEauthorblockA{\IEEEauthorrefmark{1}Laboratoire d'Informatique et de Math\'ematiques}
\IEEEauthorblockA{\IEEEauthorrefmark{2}Laboratoire d'Energ\'etique, d'Electronique et Proc\'ed\'es}
\IEEEauthorblockA{University of Reunion Island, 15 Rue Ren\'e Cassin, 97490 Sainte Clotilde, France\\email: firstname.lastname@univ-reunion.fr}}

\begin{document}

\maketitle
\begin{abstract}

Traceroute is the main tools to explore Internet path. It provides limited information about each node along the path. However, Traceroute cannot go further in statistics analysis, or \emph{Man-Machine Interface (MMI)}.  

Indeed, there are no graphical tool that is able to draw all paths used by IP routes. 
We present a new tool that can handle more than 1,000 Traceroute results, map them, identify graphically MPLS links, get information of usage of all routes (in percent) to improve the knowledge between countries' links. rTraceroute want to go deeper in usage of atomic traces. In this paper, we will discuss the concept of rTraceroute and present some example of usage.

\end{abstract}

\section{Introduction}
\label{sec:intro}
\subsection{Context}
Since the beginning of the Internet, the network has evolved from a simple graph with few vertices to a wide and a complex graph with uncountable number of nodes. 

To improve our understanding of the Internet routing and its topology, active metrology is the only way possible with the injection of packets in the network. Traceroute is one of the most widely used network measurement tools. It reports an IP address for each network-layer device along the path from a source to a destination host in an IP network. Network operators and researchers rely on Traceroute to diagnose network problems
and to infer properties of IP networks, such as the topology
of the Internet.

\begin{figure}[ht!]
\centering
\includegraphics[width=0.48\textwidth]{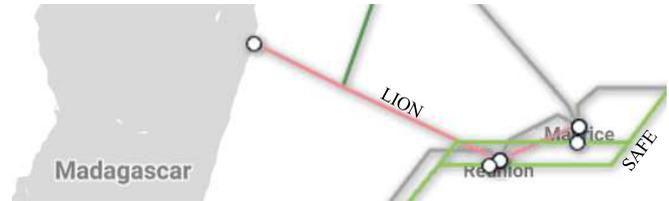}
\caption{Mascarene islands submarine cable - Source: www.submarinecablemap.com}
\label{map:cable}
\end{figure}

Reunion Island is a small territory located in the Indian Ocean, near South Africa, off the eastern coast of Madagascar. However it is also one of the overseas departments of France.
For Reunion Island, this issue is more important than other places due to its particular connectivity to the Internet based on two submarine cables, called \emph{South Africa Far-East (SAFE)} and \emph{Lower Indian Ocean Network (LION) II}, respectively in green and in red in figure~\ref{map:cable}. If in one way, we know the physical path of the Internet connection, in the other way, no one knows the logical path of our connection. This is why we want to graph the links between Reunion Island and the world-wide destination. This new representation will improve the global knowledge of the local topology. 

\subsection{Project}
\label{subsec:project}
The council of Region Reunion have financed a metrology project, with \emph{European Regional Development Funds (ERDF)} to identify the difficulties of the Reunion Island connectivity. The requirements for this project are: 

\begin{enumerate}
\item To identify the first hop after any connection leave Reunion Island and the last hop before it reaches us. This will help on the characterization of the logical exits and entries for Reunion Island.
\item To detect the \emph{MultiProtocol Label Switch (MPLS)} link between Reunion Island and the different hops. It will help to understand how are the exits and entries for Reunion Island. 
\item To focus on the regional peering problem. This item will allow us to see if the physical cable and the routing policies could be superposed.
\item To identify the most encountered node after (resp. before) leaving (resp. reaching) Reunion Island. This could improve our knowledge of the routing rules of all \emph{Internet Service Providers (ISP)} present on the territory.

\item To measure the minimal delay associated with each links between Reunion Island and the next (resp. previous) hop. It could detect prioritized link based on the final (resp. original) destination (resp. source).

\end{enumerate}
However identifying these points are not easily obtained with small data-set. We need to analyze quite big data-set, more than one probe for each \textit{ISP} present on the Island. Usually, each network tool generates its own measurements before creating an output in text and/or image format. It is difficult to combine data from several sources (resp. destination) from (resp. to) a same country without a platform measurement.

Moreover, before running an analysis we need to proceed methodically as follows:

\begin{enumerate}
\item Cleaning data-sets. 
\item Matching data-sets from different tools, with different standards. 
\item Adding Geo-localization.
\item Showing result on a map. 
\end{enumerate}

To help solve this problem, we propose \emph{rTraceroute}, a tool which is "easy to use" but can provide a map together with statistics by analyzing data from different Traceroute sources.
This tool take only one parameter: a folder which contains data from Traceroute tools. It gives us then two output: the first one is a statistic file about each node meet and the second is a map with the different link between countries. 

Thus, rTraceroute allows us to save time and to focus on the analyzing of outputs. 
The rest of this paper is organized as follows. At first, we will describe the some existing tools in section~\ref{sec:tools}.
The section~\ref{sec:tool} describes rTraceroute. An example of how it works is presented in~\ref{sec:cases} and it will be followed by a conclusion in section~\ref{sec:conclusion}.

\section{Existing tools}
\label{sec:tools}
As things stand today, no tool is available to fill our needs. However, we can find some tools that could answer some of them. Indeed, there are tools for the identification of the path from a probe to a destination. This is the reason why we have created \emph{rTraceroute} to fill the need from this project.

In this section, we will present several tools that could help us. We have categorized them as follows:
\begin{enumerate}
\item Tools that give raw results, without any statistics or map.
\item Platforms measurement that can take advantage of the previous tools and use them together.
\item Mapping tools that can draw their result on a map.
\end{enumerate}

\subsection{Raw Results}
These tools would provide us data from one measurement, but they have not been built to generate a large raw data without a script.

\subsubsection{Traceroute}
Traceroute~\cite{rfc1393} is the most popular tool to measure path length and to determine the routes that packets follow from a source node to a destination node. With the growth of the Internet, many paths can be found with Traceroute. But the main flaw of this tool is that it could only generate raw data (without any statistic or mapping).\\

\subsubsection{Paris-Traceroute}
Paris-Traceroute~\cite{augustin2006} is a Traceroute-like tool, but it is less sensitive to the load-balancing phenomena. 
The output of this tool announces some errors in the paths, like \emph{Unreachable Network (!N)} or \emph{Unreachable Host (!H)}. The using of \emph{explicit MultiProtocol Label Switch (MPLS)}~\cite{Donnet2012} link is also provided by the tool. Using Paris-Traceroute in our work could only identify the MPLS link but it has the same flaws as Traceroute.\\

\subsubsection{TraIXceroute}

TraIXceroute~\cite{Nomikos2016} is also a Traceroute-like tool with the discovering of \emph{Internet eXchange Point (IXP)} in the paths. This version of Traceroute combines two databases, \emph{PeeringDB}~\cite{peeringdb} and \emph{Packet Clearing House}~\cite{PCH}, for the identification of these particular nodes. This information could help for the reconstruction of the country crossed. Despite an accuracy of 92-93\%, this tool is not adapted for our study because it can only detect the regional routing problem with the IXP databases~\cite{Nomikos2016}.

\subsubsection{Reverse-Traceroute}
Asymetric path is frequent in the Internet, and it's very difficult to identify the reverse path. To avoid this,~\cite{katz2010} has presented \emph{Reverse-Traceroute}. This tool aims to identify the return path of a classic Traceroute, with an accuracy of 87\% for hops identification. It's mainly based on two IP options, \emph{IP Record-Route} and \emph{IP Timestamp option}. Like the original Traceroute, reverse-Traceroute can only generate Traceroute raw data but for the return way. Another method to obtain the reverse path is using a measurement platform.
\subsection{Measurement Platform}

According to~\cite{Bajpai2015survey}, an Internet measurement platform is an infrastructure of dedicated probes that periodically run network measurement tests on the Internet.
\subsubsection{Atlas RIPE NCC}
Atlas RIPE NCC~\cite{ripe2010} is an active measurement platform, which allowed to use several active tools, like \emph{ping, DNS, HTTP} or \emph{NTP} for example. There are two different tools for path measurement. The default one is Paris-Traceroute. The second one is the original Traceroute. Atlas platform return the results in \emph{JavaScript Object Notation (JSON)} format. It is possible also to map the path with \emph{OpenIPMap}~\cite{aben2015}. Despite the fact Atlas could use Paris-Traceroute, made mapping with geo-localization and compare large data produced by its measurement, this platform is not adapted to our project because it lacks the statistics part.\\

\subsubsection{Planet-Lab}
In 2003, some American researchers deployed a test-bed platform called Planet-Lab~\cite{Chun2003}.
In 2008, a European portion of Planet-Lab~\cite{EPlanetLab} have been created. 
On Planet-Lab nodes,  researchers can install the tool they need for their works. With this politic, this platform could combine several tools. Nevertheless, it is unable to identify the last (resp. first) hop after (resp. before) reaching (resp. leaving) Reunion Island. Moreover, the management of large data is not allowed on PlanetLab.\\ 

\subsubsection{Archipelago}
\emph{Center for Applied Internet Data Analysis (CAIDA)} has deployed its own platform in 2007, called Archipelago~\cite{archipelago}. The probe connected to this platform allows us to make five main measurements, including \emph{ping} and \emph{Traceroute}. But Archipelago, as Traceroute, could only generate raw data without any analysis.

\subsection{Mapped results}
\subsubsection{GTrace}
The first tool created to graph Traceroute data was GTrace~\cite{periakaruppan1999}. This tool generates its own Traceroute data and draw it on a map. It works with city name abbreviations or airport codes, lookup client and two IP databases to validate each IP addresses location. This tool was not maintained for a long time and doesn't support the x86 architecture. Due to this, we were unable to test it. 

\subsubsection{Topology Visualization Tool}
In~\cite{yang2016}, a tool centralized on analysis topology in Africa was presented. This tool can generate its own Traceroute measurement and graph it after requesting information about nodes from several databases. The design is largely based on existing visualization tools, such as~\emph{OpenIPMap project}, and focuses only on the African continent. The source code of the tool was not indicated in the article, thus no test has been made.\\

The table~\ref{tab:tools} resumes the possibilities of each tool presented previously. The cells in gray represent the available options for each tool, when the white cells mean that the option is not present. 

\begin{table*}[ht!]
\centering
\caption{Tools option availability}
\label{tab:tools}
\resizebox{\textwidth}{!}{ %
\begin{tabular}{|c|c|c|c|c|c|c|c|c|}
\hline
\textbf{Need} &  &  &  &  &  &  &  &  \\ \cline{1-1}
\textit{Tools} & \multirow{-2}{*}{\textbf{Last Hop identification}} & \multirow{-2}{*}{\textbf{MPLS Mark}} & \multirow{-2}{*}{\textbf{Statistics}} & \multirow{-2}{*}{\textbf{Geo-localization}} & \multirow{-2}{*}{\textbf{Mapping}} & \multirow{-2}{*}{\textbf{Database}} & \multirow{-2}{*}{\textbf{Regional Routing problem}} & \multirow{-2}{*}{\textbf{Large data}} \\ \hline
\textit{Traceroute} &  &  &  &  &  &  &  &  \\
\hline
\textit{Paris-Traceroute} &  & \cellcolor[HTML]{9B9B9B}YES &  &  &  &  &  &  \\ 
\hline
\textit{TraIXroute} &  &  &  &  &  & \cellcolor[HTML]{9B9B9B}YES & \cellcolor[HTML]{9B9B9B}YES &  \\
\hline
\textit{Reverse-Traceroute} &  &  &  &  &  &  &  &  \\ 
\hline
\textit{Atlas RIPE NCC} & \cellcolor[HTML]{9B9B9B}YES & \cellcolor[HTML]{9B9B9B}YES &  & \cellcolor[HTML]{9B9B9B}YES & \cellcolor[HTML]{9B9B9B}YES & \cellcolor[HTML]{9B9B9B}YES &  & \cellcolor[HTML]{9B9B9B}YES \\
\hline
\textit{PlanetLab} & \cellcolor[HTML]{9B9B9B}YES & \cellcolor[HTML]{9B9B9B}YES &  &  &  & \cellcolor[HTML]{9B9B9B}YES & \cellcolor[HTML]{9B9B9B}YES &  \\ 
\hline
\textit{Archipelago} &  &  &  &  &  &  &  &  \\ 
\hline
\textit{Gtrace} & \cellcolor[HTML]{9B9B9B}YES &  &  & \cellcolor[HTML]{9B9B9B}YES & \cellcolor[HTML]{9B9B9B}YES & \cellcolor[HTML]{9B9B9B}YES &  &  \\ 
\hline
\textit{African Visual Route} & \cellcolor[HTML]{9B9B9B}YES &  &  & \cellcolor[HTML]{9B9B9B}YES & \cellcolor[HTML]{9B9B9B}YES & \cellcolor[HTML]{9B9B9B}YES &  &  \\ 
\hline
\textit{rTraceroute} & \cellcolor[HTML]{9B9B9B}YES & \cellcolor[HTML]{9B9B9B}YES & \cellcolor[HTML]{9B9B9B}YES & \cellcolor[HTML]{9B9B9B}YES & \cellcolor[HTML]{9B9B9B}YES & \cellcolor[HTML]{9B9B9B}YES & \cellcolor[HTML]{9B9B9B}YES & \cellcolor[HTML]{9B9B9B}YES \\ 
\hline
\end{tabular}%
}
\end{table*}

The header line of the table show our needs:
\begin{itemize}
\item The identification of the last hop corresponds to the country where the node is hosted before leaving (resp. joining) a specific country. Only a graph tool can identify this point.
\item MPLS mark detect the~\emph{explicit MPLS} link between two nodes. Only Paris-Traceroute can detect these, when our rTraceroute with the graph part can detect~\emph{invisible MPLS} route.
\item Statistics part needs a large amount of data for the analyze. In general, the raw results generator does not provide any analysis. Even graphic tools can't do it because it redraws on the map on each new test. rTraceroute can analyze a large dataset and make some statistics one each meeting node.
\item The geo-localization of each IP address is based on the declaration of the RIR plus the ISP. It's well known that the Geo-localization is not accurate science. rTraceroute propose a database which could be improved by the user analysis.
\item In the general case, mapping tools need to generate their own data before draw them on a map. The first problem concerns the choice of the map and the possibility to download the final result. With rTraceroute, you can choose your own map as input and download the results for any publication or presentation. Mapping data allows us to identify the different countries pass through and detect the potential errors. With rTraceroute, these phenomena could be easily detected by the mapping and the statistics part.
\item The database is mainly used for Geo-localization of IP address or detection of the IXP.  Our database is, for now, focus on Reunion Island with the help of delay analysis.

\item The regional routing problem means the identification of peering mistakes, like boomerang-routing phenomenon~\cite{obar2012}.
\item Each tool provide to generate its own raw data one by one. rTraceroute propose to analyze a large dataset for the statistics and mapping part.

\end{itemize}

Due to the imperfections of these tools (relatively to our need), we have made the choice to develop a new tool: \emph{rTraceroute}.

\section{\lowercase{r}T\lowercase{raceroute}}
\label{sec:tool}

\subsection{The concept}
\label{subsec:concept}
rTraceroute have been created to simplify the Traceroute data file analysis. 
\begin{figure*}[!ht]
\centering

\includegraphics[width=0.98\textwidth]{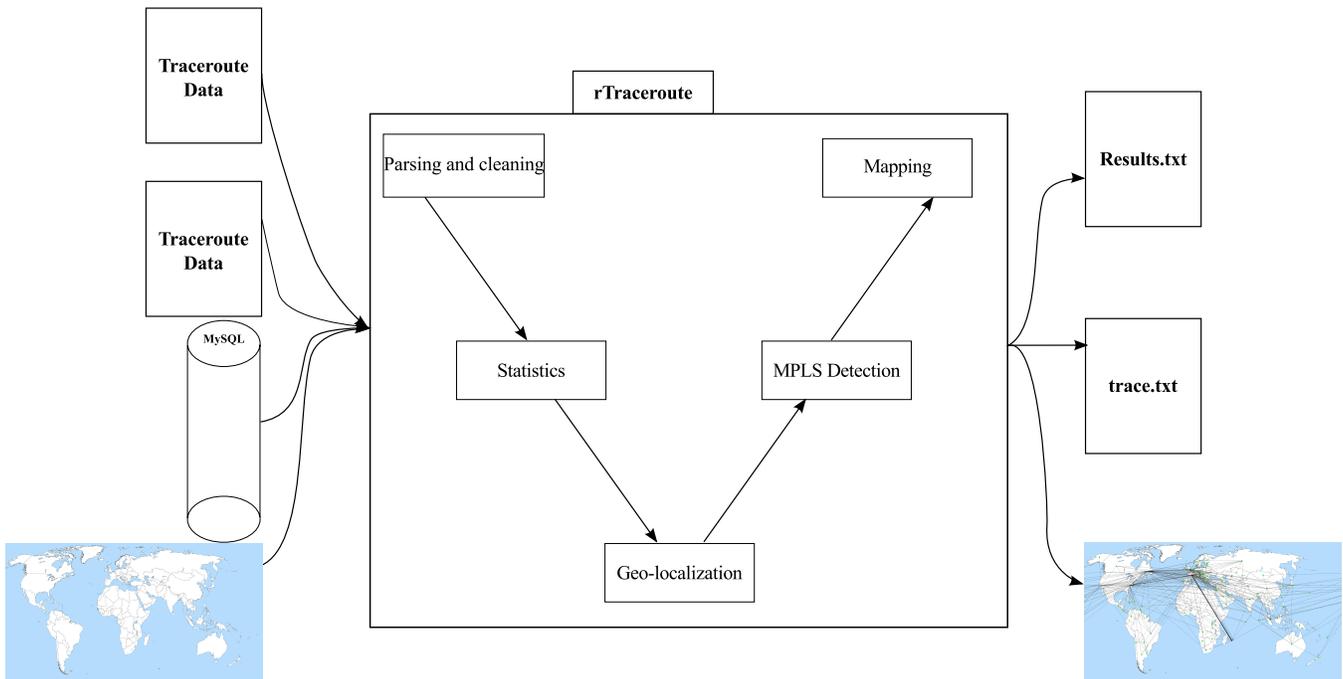}
\caption{Concept of rTraceroute}
\label{fig:concept}
\end{figure*}

On the one hand, we must process a huge number (more than 1 million) of Traceroute files. On the other hand, we are looking for performance.
The choice to handle this is to write the program with standard C, keeping in mind that it should be compatible (\emph{Linux, OS X}, ...), without any platform-specific code.
We also want the program to be easy to compile (this is why the code is a monolithic bloc, for less than three thousand lines), with a very simple Makefile.
Today, the executable run on a dedicated server with \emph{Linux Ubuntu 14.0.4 LTS}.\\

The figure~\ref{fig:concept} schematize the concept of rTraceroute. The inputs used by rTraceroute are :
\begin{itemize}
\item Several files of Traceroute data, whether it comes from Paris-Traceroute~\cite{augustin2006} or Atlas platform~\cite{ripe2010}, in text format.
\item The URI of the database used for geo-localization.
\item A Map.
\end{itemize}

Then, rTraceroute will proceed as follows:
\begin{enumerate}
\item \textbf{Parsing and Cleaning}: As the input data from Paris-Traceroute~\cite{augustin2006} or Atlas platform~\cite{ripe2010} are text files, the program must parse JSON files as well as plain text files.
The parse procedure for JSON files is made with NXJson~\cite{NXJSON}.
To be more efficient, the program has a "cleaning procedure" (as it has been explained in~\ref{subsec:project}
) that eliminates useless (or corrupted) files. The pre-parsing subroutine uses the following exclusion pattern:
\begin{itemize}
\item JSON corrupted.
\item JSON file contains 3 last RTT as \texttt{0} (zero) [unreachable].
\item Traceroute corrupted [file contains more than one Traceroute].
\item Traceroute contains 3 stars on the last hop [host unreachable].
\item Traceroute contains \texttt{!H} or \texttt{!N} or \texttt{WARN} string [unreachable].
\end{itemize}
\item \textbf{Statistics}: During parsing files, rTraceroute create an internal linked-chain memorizing all information. Each IP address and hop are marked with an on the fly calculated RTT; later completed with the localization (country) from the MySQL internal database.
\item \textbf{Geo-localization}: 
From this step, we need to geo-localize IP address. This is why the program needs a MySQL database (connection information is "hard-coded" in main.h).
.
Two tables are used: the first one to geo-localize IP address and the second one to place the country on a map.\\
Then, to be able to draw each trace, \textit{gdlib} has been used.
\\
To store all information obtained during the process, three kinds of tuples are used:
\begin{itemize}
\item \texttt{\{hop, ip address, rtt, occurrences\}} is added when reading files.
\item \texttt{\{mapx1, mapy1, mapx2, mapy2, ip\_address1, ip\_address2, link\}} is created when drawing is done, to handle all MPLS links.
\item \texttt{\{xdeb, ydeb, xfin, yfin, occurences, rtt, mpls\}} is also handled when the program draws.
\end{itemize}
\item \textbf{MPLS Detection}:~\emph{Paris-Traceroute} can detect~\emph{explicit MPLS} links and marked them in its raw data. When rTraceroute find the marks, it memorized the link to plot them in a different coloring at the end of the mapping.

\item \textbf{Mapping}: Some useful options have been implemented as the program can handle html color's name and create statistics:
\begin{itemize}
\item It can draw only the last links (the last hop before coming in one country).
\item It can make the thickness of the link proportional of its use (with the \texttt{-redraw} option).
\item It colorize MPLS links.
\end{itemize}
\end{enumerate}

The program produces three result files:
\begin{itemize}

\item A simple one that contains only 5 columns:  \texttt{hop\_position, IP address, occurrence, average\_delay, country}, used for a later statistical purpose.

\item A very verbose one called  \texttt{trace.txt} that explicitly details all internal operations. From this text file, it is possible to extract and produce more results (as statistics in percent of use's link).
for examples:
\begin{itemize}
\item to extract all MPLS links: \newline
\texttt{ \$ grep 'lien.*MPLS' trace.txt } \newline
and we obtain this kind of result: \newline
\texttt{ // lien <-> MPLS: 93.17.132.110 109.24.74.178 }
\item to find information on each drawing line (x1,y1 to x2,y2), occurrences, occurrence in percent between all lines, minimal time, geo-localization, in text, for all segments:\newline
\texttt{ grep '\_\_ trajet' trace.txt }\newline
and we obtain this kind of result: \newline
\texttt{ \_\_ trajet: 3626 1638 -> 2581 1582: 105 (2.81\%) [AU - RE et temps min 505.39 (2.81\%)] }
\end{itemize}

\item Two graphical files (maps): all segments of the input files on the first one and only the segments of the last hop for a destination point (x,y) on the last map.
\end{itemize}

rTraceroute is the most adapted tool for our project. Indeed, its capacity of mapping help us to answer the first need of the project, which is the identification of the first (resp. lat) hop after (resp. before) leave (resp. reach) Reunion Island. The map function also help us on the third item of the project: the regional peering problem. Paris-Traceroute allow to detect \emph{explicit MPLS} link. With rTraceroute, even the \emph{invisible MPLS}~\cite{Donnet2012} link is shown with the maps option. The statistics part of rTraceroute permits to answer the last two points, which are respectively the identification of the most encountered node after (resp. before) leaving (resp. reaching) Reunion Island and the minimal delay associated with each links between Reunion Island and the next (resp. previous) hop.

\section{Application of rTraceroute on Reunion Island}
\label{sec:cases}

This paper has begun with a quick presentation of different active tools or platforms. This part will show how rTraceroute used the data provided by these tools for analysis path from a country to several destinations. The section~\ref{subsec:from} explain how rTraceroute can be used with direct Paris-Traceroute data, when~\ref{subsec:toward} used Paris-Traceroute JSON formatted from Atlas RIPE NCC platform as data. The results have been previously analyzed in~\cite{Noordally2016} and are available on the website of the authors.

\subsection{Paris-Traceroute dataset}
\label{subsec:from}

\begin{figure*}[ht!]
 \centering
 \includegraphics[width=0.98\textwidth]{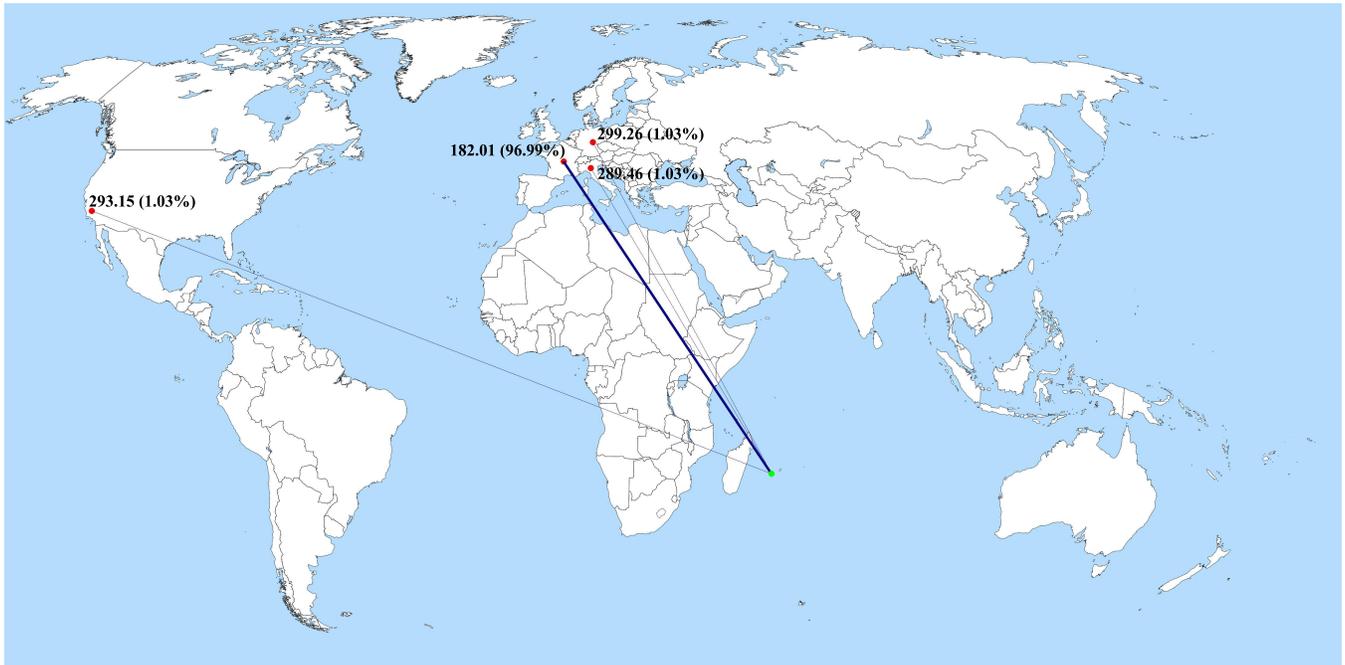}
 \caption{Paris-Traceroute results example}
 \label{fig:rpi}
\end{figure*}

In this example, we will test rTraceroute on the data-set publicly available and coming from~\cite{Noordally2016}. This large data was obtained during one month, between third of July and third of August 2016. With 27 Raspberry Pi\cite{raspberry_maksimovic2014} used as measurement probes, each day we have tried to reach $10,000$ IPv4 spread around the world as destination. At the end of the measurement campaign we have obtained a total of $6,458,025$ raw traces.\\
Combined with rTraceroute, the raw data have been reduced to $1,015,180$ available for path analysis. The using of rTraceroute on Paris-Traceroute data have these advantages:
\begin{itemize}
\item The first one concern the raw data. The tools can filter only the data which fill the selection criteria presented in the section~\ref{sec:tool} 
\item Mapping the path makes the path analysis between countries easier. Along with mapping, it is possible to print only the first (resp. last) node before (resp. after) leaving (resp. reaching) a country with the path and the delay associated.
\item Lastly, rTraceroute can change all \emph{explicit MPLS} link's color found by Paris-Traceroute. The mapping would permit the detection of \emph{invisible MPLS} links.
\end{itemize}
The map~\ref{fig:rpi} represent the results for the first hop when data leave Reunion Island.\\
In this figure, we can see the first hop after Reunion Island. Despite the presence of two physical connections with Asia and Africa~\ref{map:cable}, we can notice that most of the first hops are not connected to these continents but to Europe.

Another interesting point is the presence of a direct link between Reunion and USA, sign of~\emph{invisible MPLS} link. We can also see that more of 96\% of our data going through France before joining the destination, even if the destination is close to Reunion Island. 
A detailed analysis of the results are available in~\cite{Noordally2016}.\\
The analysis of the same data with only a \emph{BASH} script has taken more than one day. With rTraceroute, around 10 minutes is sufficient.

\subsection{Atlas RIPE NCC dataset}
\label{subsec:toward}

\begin{figure*}[ht!]
\centering
\includegraphics[width=0.98\textwidth]{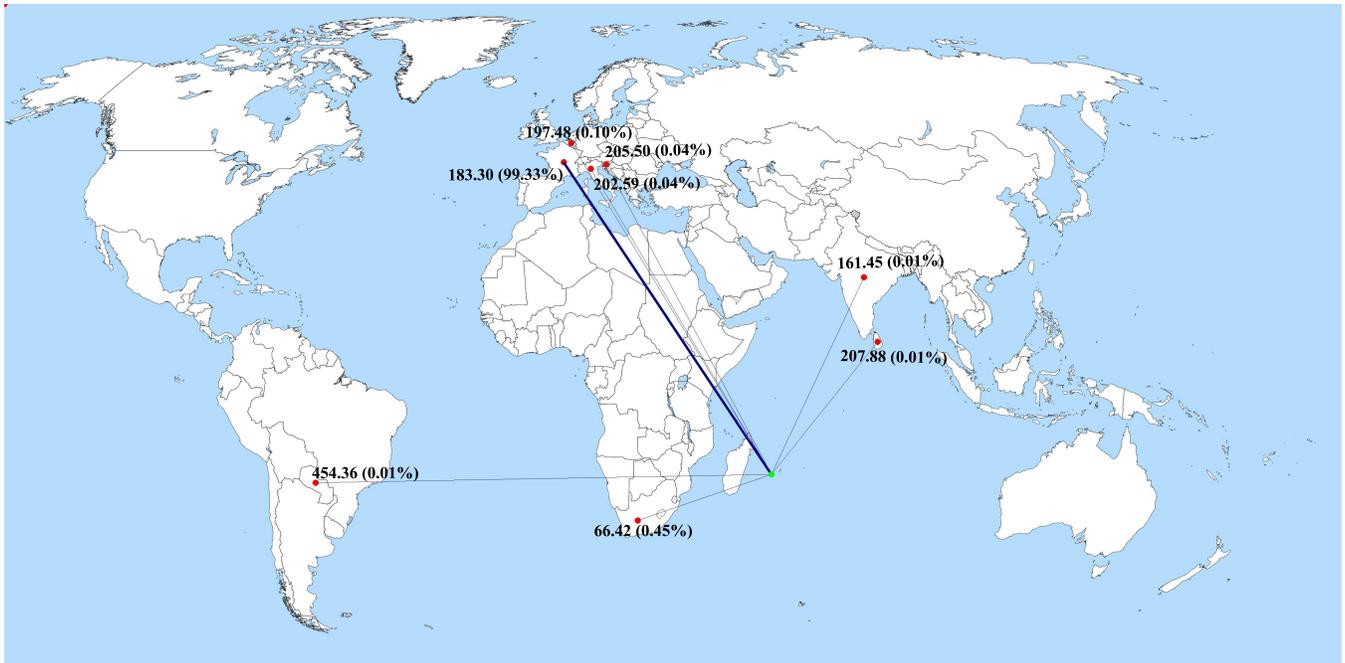}
\caption{Example of results from rTraceroute used in combination with Atlas}
\label{fig:atlas_all}
\end{figure*}

In this case study, we would analysis $300,000$ measurement towards Reunion Island from all world. Our trace includes measurement performed from the third of July to the third of August 2016. $1,000$ atlas probes would reach ten Raspberry Pi~\cite{raspberrypi} deployed over Reunion Island. The atlas probe was selected with the same distribution as the IPv4 addresses for the Paris-Traceroute case.\\
Without rTraceroute, we would have been constraint by using OpenIPMap~\cite{aben2015}. The tools present some limits: it requires to add manually the other path measurements to compare the results and the maps can't be downloaded. 
The "cleaning step" of rTraceroute have reduced our data to $38,714$ traces.
The advantage of combine rTraceroute with Atlas is the same as with the other Traceroute tools. But in this context, we can highlight three benefits:
\begin{itemize}
\item First we can easily combine data from several Atlas math measurements for the analysis. A script could download the JSON file and add it in a folder which could also be analyzed by rTraceroute.
\item The maps are hosted on the system which run rTraceroute. Maps can be downloaded and used after.
\item Eventually, the statistics file created by rTraceroute is very important. Each line of the file contains the position of an IP address meet during the measurement, the IP address, the occurrence of this couple, the meaning delay and the country associated with the IP address.
\end{itemize}
The map~\ref{fig:atlas_all} is an example of all path existing to join Reunion Island using Atlas and rTraceroute.
On the figure~\ref{fig:atlas_all}, we can see a direct link between Paraguay and Reunion Island without a submarine cable existing. This proves us that an \emph{invisible MPLS} link exists. Without rTraceroute, this analysis will be more difficult to be found. We can also notice the use of the section of the submarine cable from Asia and Africa. The percentage of using the exit point for submarine cable in the two continent is less than 0.5\%, when more than 99\% going through France, with a minimal delay of $183.30$ms.\\
This campaign of measurement from the world towards Reunion Island has provided $300,000$ JSON raw data for a time analysis of 3 minutes with rTraceroute. A \emph{BASH} script developed for the analysis of these raw data has taken more than half a day.

\section{Conclusion}
\label{sec:conclusion}

Studying the Internet connectivity of Reunion Island is important for the economic development of the island. An~\emph{ERDF} project has been funded to identify the difficulties of the Internet connectivity of the Island.
Several tools are available, but none of them could cover our needs. Nearly no tools can handle a large number of Traceroute-files and draw all the path on a map. This why we have implemented a new tool, called~\emph{rTraceroute}.
It also offers us a new approach to observe the Internet topology with a different point of view.
As for today, there is no IP geo-localization database reliable enough to plot every IP found in the traces. The next step will be to improve IP geo-localization of each point, based on delay between to point and a system of triangulation.
IPv6 deployment is in its infancy: the routes could not be optimized. Adding IPv6 in rTraceroute could also help the deployment of this new protocol in the area, but some adaptations need to be done for IPv6 because the actual work is based on IPv4.
Exportation of maps in \emph{Encapsulated PostScript (EPS)} or \emph{Scalable Vector Graphic (SVG)} format is also planned in the future (actually, the maps result are only in PNG format).

\bibliography{metrology}

\begin{thebibliography}{10}
\providecommand{\url}[1]{#1}
\csname url@samestyle\endcsname
\providecommand{\newblock}{\relax}
\providecommand{\bibinfo}[2]{#2}
\providecommand{\BIBentrySTDinterwordspacing}{\spaceskip=0pt\relax}
\providecommand{\BIBentryALTinterwordstretchfactor}{4}
\providecommand{\BIBentryALTinterwordspacing}{\spaceskip=\fontdimen2\font plus
\BIBentryALTinterwordstretchfactor\fontdimen3\font minus
  \fontdimen4\font\relax}
\providecommand{\BIBforeignlanguage}[2]{{%
\expandafter\ifx\csname l@#1\endcsname\relax
\typeout{** WARNING: IEEEtran.bst: No hyphenation pattern has been}%
\typeout{** loaded for the language `#1'. Using the pattern for}%
\typeout{** the default language instead.}%
\else
\language=\csname l@#1\endcsname
\fi
#2}}
\providecommand{\BIBdecl}{\relax}
\BIBdecl

\bibitem{rfc1393}
G.~S. Malkin, ``Traceroute using an {IP} option,'' 1993.

\bibitem{augustin2006}
B.~Augustin, X.~Cuvellier, B.~Orgogozo, F.~Viger, T.~Friedman, M.~Latapy,
  C.~Magnien, and R.~Teixeira, ``Avoiding traceroute anomalies with paris
  traceroute,'' in \emph{Proceedings of the 6th ACM SIGCOMM conference on
  Internet measurement}.\hskip 1em plus 0.5em minus 0.4em\relax ACM, 2006, pp.
  153--158.

\bibitem{Donnet2012}
B.~Donnet, M.~Luckie, P.~M{\'e}rindol, and J.-J. Pansiot, ``Revealing mpls
  tunnels obscured from traceroute,'' \emph{ACM SIGCOMM Computer Communication
  Review}, vol.~42, no.~2, pp. 87--93, 2012.

\bibitem{Nomikos2016}
G.~Nomikos and X.~Dimitropoulos, ``traixroute: Detecting ixps in traceroute
  paths,'' in \emph{International Conference on Passive and Active Network
  Measurement}.\hskip 1em plus 0.5em minus 0.4em\relax Springer, 2016, pp.
  346--358.

\bibitem{peeringdb}
\BIBentryALTinterwordspacing
P.~DB, ``Peeringdb facilitates the exchange of information related to
  peering.'' [Online]. Available: \url{https://www.peeringdb.com/}
\BIBentrySTDinterwordspacing

\bibitem{PCH}
P.~C. House, ``Internet exchange directory,'' 2014.

\bibitem{katz2010}
E.~Katz-Bassett, H.~V. Madhyastha, V.~K. Adhikari, C.~Scott, J.~Sherry,
  P.~Van~Wesep, T.~E. Anderson, and A.~Krishnamurthy, ``Reverse traceroute.''
  in \emph{NSDI}, vol.~10, 2010, pp. 219--234.

\bibitem{Bajpai2015survey}
V.~Bajpai and J.~Sch{\"o}nw{\"a}lder, ``A survey on internet performance
  measurement platforms and related standardization efforts,'' \emph{IEEE
  Communications Surveys \& Tutorials}, vol.~17, no.~3, pp. 1313--1341, 2015.

\bibitem{ripe2010}
\BIBentryALTinterwordspacing
R.~NCC, ``{RIPE} atlas,'' 2010. [Online]. Available:
  \url{https://atlas.ripe.net}
\BIBentrySTDinterwordspacing

\bibitem{aben2015}
\BIBentryALTinterwordspacing
E.~Aben, ``Infrastructure geolocation - plan of action,'' 2015. [Online].
  Available:
  \url{https://labs.ripe.net/Members/emileaben/infrastructure-geolocation-plan-of-action}
\BIBentrySTDinterwordspacing

\bibitem{Chun2003}
\BIBentryALTinterwordspacing
B.~Chun, D.~Culler, T.~Roscoe, A.~Bavier, L.~Peterson, M.~Wawrzoniak, and
  M.~Bowman, ``{PlanetLab}: An overlay testbed for broad-coverage services,''
  \emph{SIGCOMM Comput. Commun. Rev.}, vol.~33, no.~3, pp. 3--12, Jul. 2003.
  [Online]. Available: \url{http://doi.acm.org/10.1145/956993.956995}
\BIBentrySTDinterwordspacing

\bibitem{EPlanetLab}
\BIBentryALTinterwordspacing
``{PlanetLab} europe, an open platform for developing, deploying, and accessing
  planetary-scale services.'' [Online]. Available: \url{http://planet-lab.eu/}
\BIBentrySTDinterwordspacing

\bibitem{archipelago}
K.~Claffy, Y.~Hyun, K.~Keys, M.~Fomenkov, and D.~Krioukov, ``Internet mapping:
  from art to science,'' in \emph{Conference For Homeland Security, 2009.
  CATCH'09. Cybersecurity Applications \& Technology}.\hskip 1em plus 0.5em
  minus 0.4em\relax IEEE, 2009, pp. 205--211.

\bibitem{periakaruppan1999}
R.~Periakaruppan, E.~Nemeth \emph{et~al.}, ``{GTrace}: A graphical traceroute
  tool.'' in \emph{LISA}, vol.~99, 1999, pp. 69--78.

\bibitem{yang2016}
C.~Yang, H.~Suleman, and J.~Chavula, ``A topology visualisation tool for
  national research and education networks in {Africa},'' in \emph{IST-Africa
  Week Conference, 2016}.\hskip 1em plus 0.5em minus 0.4em\relax IIMC, 2016,
  pp. 1--11.

\bibitem{obar2012}
J.~A. Obar and A.~Clement, ``Internet surveillance and boomerang routing: A
  call for {Canadian} network sovereignty,'' in \emph{TEM 2013: Proceedings of
  the Technology \& Emerging Media Track-Annual Conference of the Canadian
  Communication Association (Victoria)}, 2012.

\bibitem{NXJSON}
\BIBentryALTinterwordspacing
Y.~Stavnichiy, ``Nxjson, a light json parser written in c.'' [Online].
  Available: \url{https://bitbucket.org/yarosla/nxjson/}
\BIBentrySTDinterwordspacing

\bibitem{Noordally2016}
R.~Noordally, X.~Nicolay, P.~Anelli, R.~Lorion, and P.~U. Tournoux, ``Analysis
  of internet latency : the reunion island case,'' in \emph{Asian Internet
  Engineering Conference}.\hskip 1em plus 0.5em minus 0.4em\relax ACM, 2016,
  pp. 49--56.

\bibitem{raspberry_maksimovic2014}
M.~Maksimovi{\'c}, V.~Vujovi{\'c}, N.~Davidovi{\'c}, V.~Milo{\v{s}}evi{\'c},
  and B.~Peri{\v{s}}i{\'c}, ``Raspberry pi as internet of things hardware:
  performances and constraints,'' \emph{design issues}, vol.~3, p.~8, 2014.

\bibitem{raspberrypi}
\BIBentryALTinterwordspacing
``{Raspberry Pi}, official website.'' [Online]. Available:
  \url{http://www.raspberrypi.org/}
\BIBentrySTDinterwordspacing

\end{thebibliography}
\bibliographystyle{IEEEtran}

\appendix
A version of rTraceroute is in public access at \url{http://lim.univ-reunion.fr/rTraceroute}
\end{document}